\documentclass[final]{aipproc}

\layoutstyle{8x11single}

\newcommand{\msun}{\,\rm M_\odot}

\newcommand{\be}{\begin{equation}}
\newcommand{\ee}{\end{equation}}
\newcommand{\ba}{\begin{eqnarray}}
\newcommand{\ea}{\end{eqnarray}}

\newcommand{\msub}{M_{\rm sub}}

\newcommand{\rvcmax}{r_{\rm Vmax}}

\newcommand{\sigv}{\langle \sigma {\rm v} \rangle}

\begin{document}

\title{GLAST and Dark Matter Substructure in the Milky Way}


\author{Michael Kuhlen}{
  address={School of Natural Science, Institute for Advanced Study, Einstein Lane, Princeton, NJ 08540, USA}
}

\author{J\"urg Diemand}{
  address={Department of Astronomy and Astrophysics, UC Santa Cruz, 1156 High Street, Santa Cruz, CA, USA},
  altaddress={Hubble Fellow}
}

\author{Piero Madau}{
  address={Department of Astronomy and Astrophysics, UC Santa Cruz, 1156 High Street, Santa Cruz, CA, USA},
  altaddress={Max-Planck-Institut f\"ur Astrophysik, Karl-Schwarzschild-Str. 1, 85740 Garching, Germany}
}

\begin{abstract}
We discuss the possibility of GLAST detecting gamma-rays from the
annihilation of neutralino dark matter in the Galactic halo. We have
used ``Via Lactea'', currently the highest resolution simulation of
cold dark matter substructure, to quantify the contribution of
subhalos to the annihilation signal. We present a simulated allsky map
of the expected gamma-ray counts from dark matter annihilation,
assuming standard values of particle mass and cross section. In this
case GLAST should be able to detect the Galactic center and several
individual subhalos.
\end{abstract}

\maketitle

%%%%%%%%%%%%%%%%%%%%%%%%%%%%%%%%%%%%%%%%%%%%
%% MAINMATTER
%%%%%%%%%%%%%%%%%%%%%%%%%%%%%%%%%%%%%%%%%%%%

\section{Introduction}

One of the most exciting discoveries that the \textit{Gamma-ray Large
Area Space Telescope} (GLAST) could make, is the detection of
gamma-rays from the annihilation of dark matter (DM). Such a
measurement would directly address one of the major physics problems
of our time: the nature of the DM particle.

Whether or not GLAST will actually detect a DM annihilation signal
depends on both unknown particle physics and unknown astrophysics
theory. Particle physics uncertainties include the type of particle
(axion, neutralino, Kaluza-Klein particle, etc.), its mass, and its
interaction cross section. From the astrophysical side it appears that
DM is not smoothly distributed throughout the Galaxy halo, but instead
exhibits abundant clumpy substructure, in the form of thousands of
so-called subhalos. The observability of DM annihilation radiation
originating in Galactic DM subhalos depends on their abundance,
distribution, and internal properties. 

Numerical simulations have been used in the past to estimate the
annihilation flux from DM substructure
\citep{CalcaneoRoldan2000,Stoehr2003,Diemand2006,Diemand2007a}, but
since the subhalo properties, especially their central density
profile, which determines their annihilation luminosity, are very
sensitive to numerical resolution, it makes sense to re-examine their
contribution with higher resolution simulations.

\section{DM annihilation in Substructure}

Here we report on the substructure annihilation signal in ``Via
Lactea'', the currently highest resolution simulation of an individual
DM halo. Details about this simulation, including the properties of
the host halo and its substructure population, can be found in
\citep{Diemand2007a, Diemand2007b}. To briefly summarize: The central
halo is resolved with $\sim 200$ million high resolution DM particles,
corresponding to a particle mass of $M_p= 2 \times 10^4 \msun$. At
$z=0$ the host halo has a mass of $M_{200} = 1.8 \times 10^{12}
\msun$, and it underwent its last major merger at $z=1.7$. In total we
resolve close to 10,000 subhalos, which make up $5.3\%$ of the host
halo mass. The subhalo mass function is well approximated by a
powerlaw $dN/d\ln M \propto M^{-1}$ over three orders of magnitude
down to the resolution limit of about 200 particles per subhalo ($\sim
4 \times 10^6 \msun$). This power law slope corresponds to equal mass
in substructure per decade, and it implies that the total subhalo mass
fraction has not yet converged. Future simulations with even lower
particle masses will presumably find an even larger subhalo mass
fraction. A limitation of this present simulation is that it
completely neglects the effects of baryons. Gas cooling will likely
increase the DM density in the central regions of the host halo
through adiabatic compression \citep{Blumenthal1986}. However, because
of their shallower potential wells, the DM distribution in galactic
subhalos is unlikely to be significantly altered by baryonic effects.

\begin{center}
\begin{figure}
\includegraphics[width=0.5\textwidth]{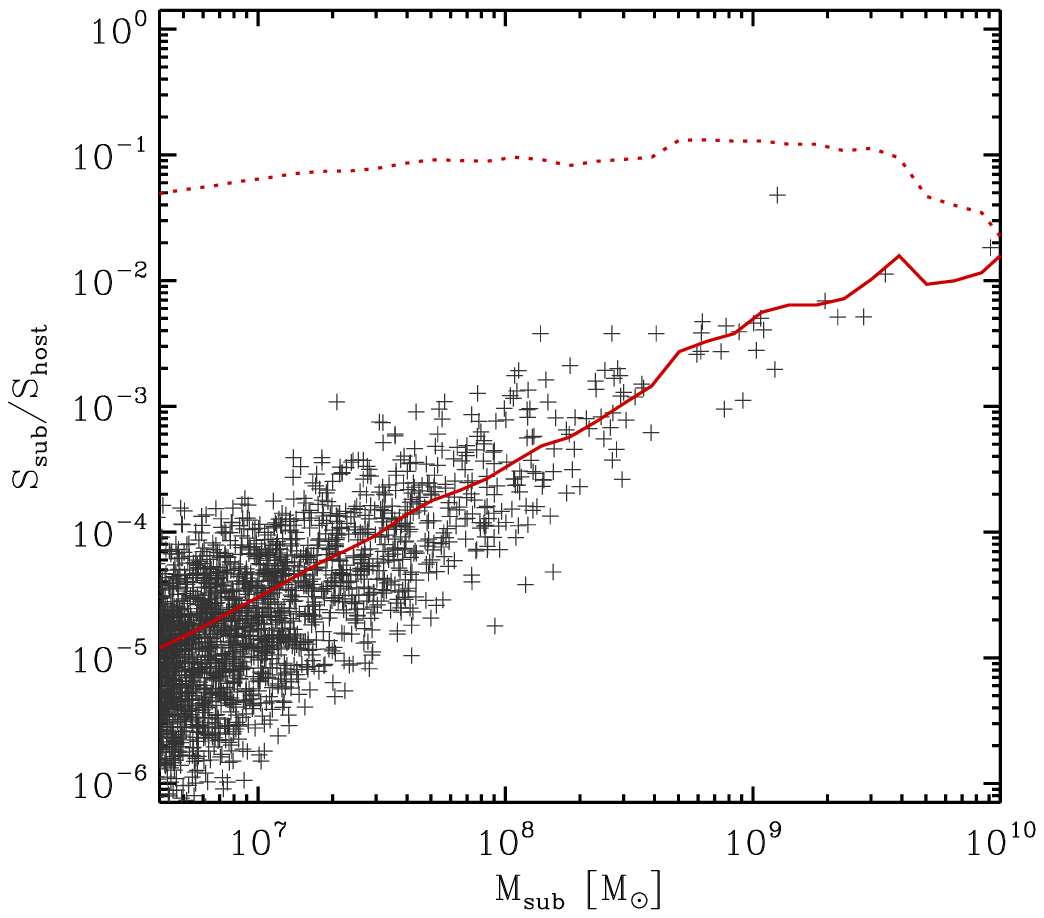}
\includegraphics[width=0.5\textwidth]{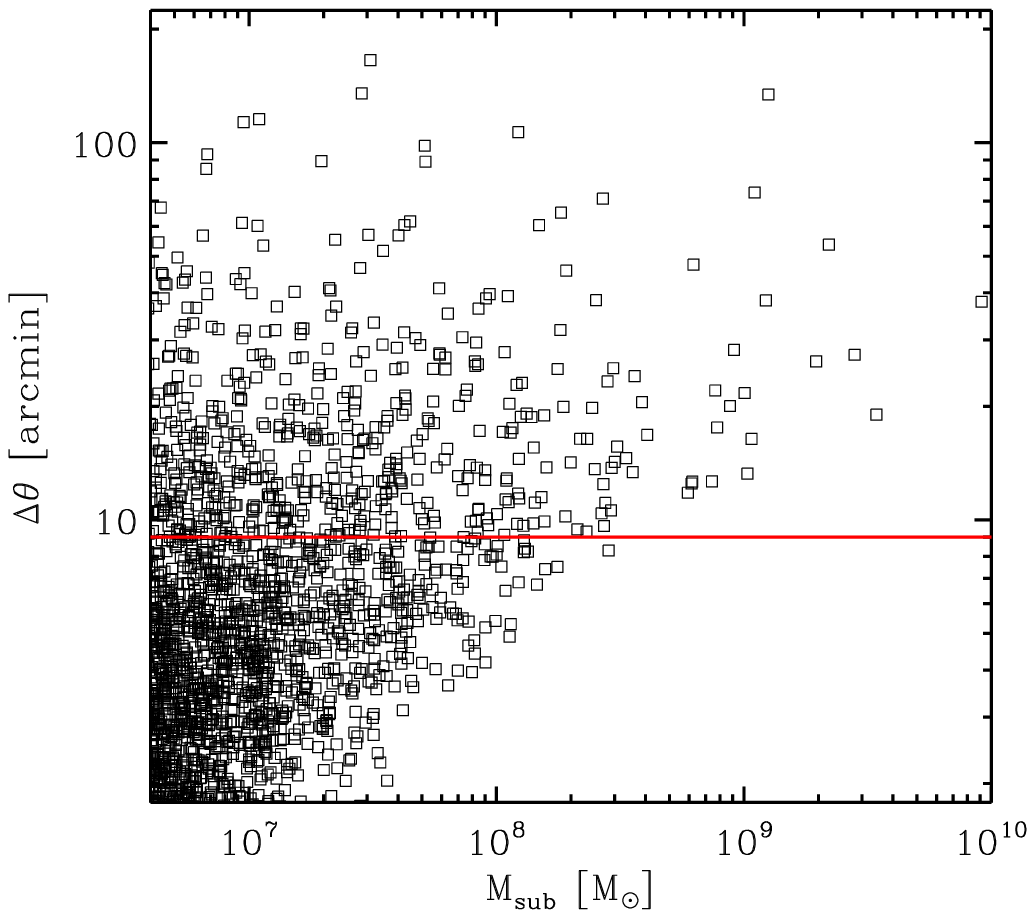}
\caption{\textit{Left panel}: The annihilation signal of individual
subhalos (\textit{crosses}) in units of the total luminosity of the
spherically averaged host halo. The curves are the average
(\textit{solid}) and total (\textit{dotted}) signal in a sliding
window over one decade in mass. \textit{Right panel}: The angular size
subtended by $2.0 r_s$ for a fiducial observer located 8 kpc from the
halo center vs. the subhalo tidal mass. For an NFW density profile
$\sim 90\%$ of the total luminosity originates within $r_s$. The
expected GLAST 68\% angular resolution at $>10$ GeV of 9 arcmin is
denoted by the solid horizontal line.}
\label{fig:annihilation_luminosity}
\end{figure}
\end{center}

We approximate the annihilation luminosity of an individual subhalo by
\begin{equation}
S_{\rm sub,i} = \int_{V_i} \rho^2_{\rm sub} dV_i = \sum_{j \epsilon \{P_i\}} \rho_j m_p,
\label{eq:annihilation_luminosity}
\end{equation}
where $\rho_j$ is the density of the $j^{\rm th}$ particle (estimated
using a 32 nearest neighbor SPH kernel), and $\{P_i\}$ is the set of
all particles belonging to halo $i$. In the left panel of
Figure~\ref{fig:annihilation_luminosity} we plot $S_{\rm sub}$
normalized by $S_{\rm host}$, the total luminosity of the spherically
averaged host halo.

We find that the subhalo luminosity is proportional to its mass. Given
our measured substructure abundance of $dN/d\ln \msub \propto
\msub^{-1}$, this implies a total subhalo annihilation luminosity that
is approximately constant per decade of substructure mass, as the
Figure shows (dotted line). We measure a total annihilation luminosity
from the host halo that is a factor of 2 higher than the
spherically-averaged smooth signal, obtained by integrating the square
of the binned radial density profile. About half of this boost is due
to resolved substructure, and we attribute the remaining half to other
deviations from spherical symmetry. Similar boost factors may apply to
the luminosity of individual subhalos as well (see next section).

The detectability of DM annihilation originating in subhalos depends
not only on their luminosity, but also on the angular size of the
sources in the sky, which we can constrain by ``observing'' the
subhalo population in our simulation. For this purpose we have picked
a fiducial observer position, located 8 kpc from the halo center along
the intermediate axis of the triaxial host halo mass distribution. In
the right panel of Figure~\ref{fig:annihilation_luminosity} we plot
the angular size $\Delta \theta$ of the subhalos for this observer
position. For an NFW density profile with scale radius $r_s$, about
90\% of the total annihilation luminosity originates within $r_s$. We
define $\Delta \theta$ to be the angle subtended by $\rvcmax/2.16$,
where $\rvcmax$ is the radius of the peak of the circular velocity
curve $V_c(r)^2 = GM(<r)/r$, which is equal to $2.16 r_s$ for an NFW
profile. GLAST's expected 68\% containment angular resolution for
photons above 10 GeV is 9 arcmin. We find that (553, 85, 20) of our
subhalos have angular sizes greater than (9, 30, 60) arcmin. In the
following section we consider the brightness of these subhalos and
discuss the possibility of actually detecting some of them with GLAST.

\section{DM annihilation allsky map}
\label{sec:allsky}

\begin{figure}
\includegraphics[width=\textwidth]{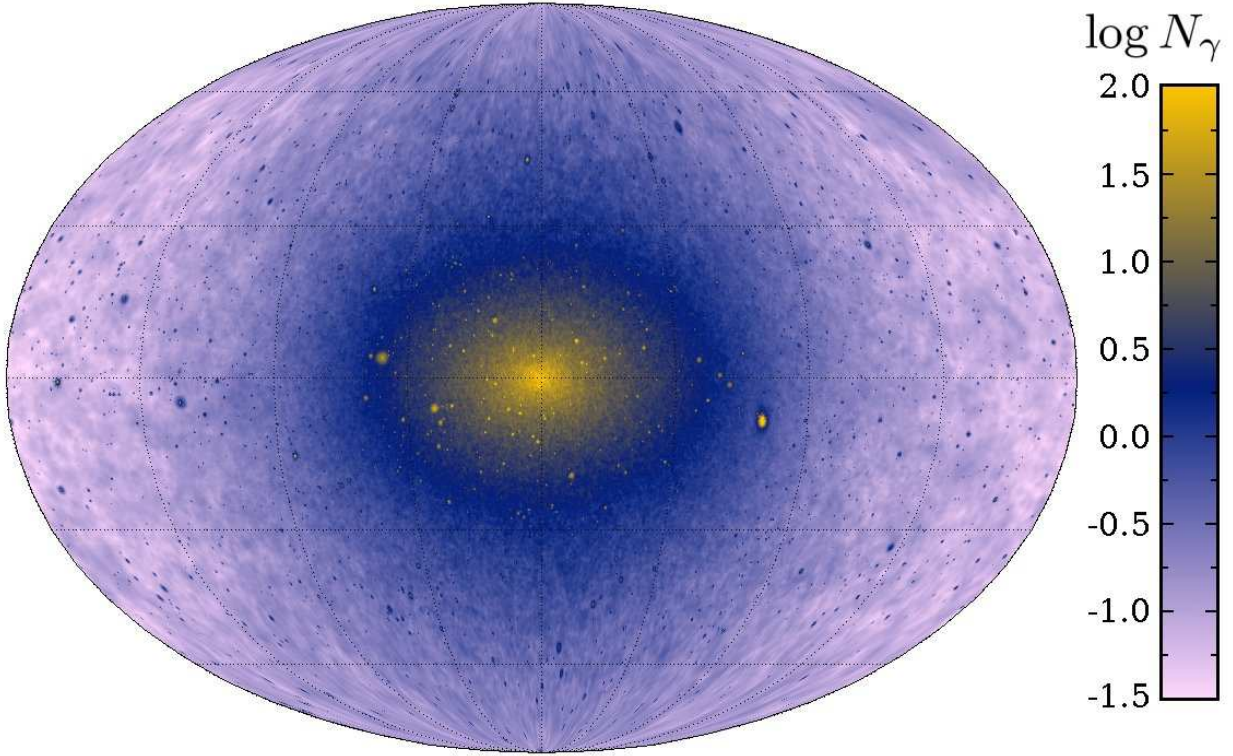}
\caption{Simulated GLAST allsky map of neutralino DM annihilation in
the Galactic halo, for a fiducial observer located 8 kpc from the halo
center along the intermediate principle axis. We assumed $M_\chi = 46$
GeV, $\sigv=5 \times 10^{-26}$ cm$^3$ s$^{-1}$, a pixel size of 9
arcmin, and a 2 year exposure time. The flux from the subhalos has
been boosted by a factor of 10 (see text for explanation). Backgrounds
and known astrophysical gamma-ray sources have not been included.}
\label{fig:allsky}
\end{figure}

Using the DM distribution in our Via Lactea simulation, we have
constructed allsky maps of the gamma-ray flux from DM annihilation in
our Galaxy. As an illustrative example we have elected to pick a
specific set of DM particle physics and realistic GLAST/LAT
parameters. This allows us to present maps of expected photon counts.

The number of detected DM annihilation gamma-ray photons from a solid
angle $\Delta \Omega$ along a given line of sight ($\theta$, $\phi$)
over an integration time of $\tau_{\rm exp}$ is given by
\begin{equation}
N_\gamma(\theta,\phi) = \Delta\Omega \; \tau_{\rm exp}
\frac{\sigv}{M_\chi^2} \left[ \int_{E_{\rm th}}^{M_\chi}
\left(\!\frac{dN_\gamma}{dE}\!\right) A_{\rm eff}(E) dE \right]
\int_{\rm los} \! \rho(l)^2 dl,
\label{eq:annihilation}
\end{equation}
where $M_\chi$ and $\sigv$ are the DM particle mass and
velocity-weighted cross section, $E_{\rm th}$ and $A_{\rm eff}(E)$ are
the detector threshold and energy-dependent effective area, and
$dN_\gamma/dE$ is the annihilation spectrum.

We assume that the DM particle is a neutralino and have chosen
standard values for the particle mass and annihilation cross section:
$M_{\chi}=46$ GeV and $\sigv=5 \times 10^{-26}$ cm$^3$ s$^{-1}$.
These values are somewhat favorable, but well within the range of
theoretically and observationally allowed models. As a caveat we note
that the allowed $M_\chi$-$\sigv$ parameter space is enormous (see
e.g. \citep{Colafrancesco2006}), and it is quite possible that the true
values lie orders of magnitude away from the chosen ones, or indeed
that the DM particle is not a neutralino, or not even weakly
interacting at all. We include only the continuum emission due to the
hadronization and decay of the annihilation products ($b\bar{b}$ and
$u\bar{u}$ only, for our low $M_\chi$) and use the spectrum
$dN_\gamma/dE$ given in \citep{Bergstroem1998}.

For the detector parameters we chose an exposure time of $\tau_{\rm
exp}=2$ years and a pixel angular size of $\Delta \theta=9$ arcmin,
corresponding to the 68\% containment GLAST/LAT angular
resolution. For the effective area we used the curve published on the
GLAST/LAT performance website \citep{glast} and adopted a threshold
energy of $E_{\rm th}=0.45$ GeV (chosen to maximize the significance,
see below). The fiducial observer is located 8 kpc from the center
along the intermediate principle axis of the host halo's ellipsoidal
mass distribution.

Lastly, we applied a boost factor of 10 to all subhalo fluxes. The
motivation for this boost factor is twofold: First, we expect the
central regions of our simulated subhalos to be artificially heated
due to numerical relaxation, and hence less dense and less luminous
than in reality. Secondly, we expect the subhalo signal to be boosted
by its own substructure. We in fact observe sub-subhalos in the most
massive of Via Lactea's subhalos \citep{Diemand2007a}, and this
sub-substructure, and indeed sub-sub-substructure, etc., will lead to
a boost in the annihilation luminosity analogous to the one for the
whole host halo, discussed in the previous section. An analytical
model \citep{Strigari2006} for subhalo flux boost factors gives boosts
from a few up to $\sim 100$, depending on the slope and lower mass
cutoff of the subhalo mass function.

Figure~\ref{fig:allsky} shows the resulting allsky map in a Mollweide
projection. The coordinate system has been rotated such that the major
axis of the host halo ellipsoid is aligned with the horizontal
direction, which would also correspond to the plane of the Milky Way
disk, if its angular momentum vector were aligned with the minor axis
of the host halo. The halo center (at $l=0^\circ$, $b=0^\circ$) is the
brightest source of annihilation radiation, but the most massive
subhalo (at around $l=+70^\circ$, $b=-10^\circ$) is of comparable
brightness. Additionally a large number of individual subhalos are
clearly visible, especially towards the halo center ($-90^\circ < l <
+90^\circ$, $-60^\circ < b < +60^\circ$).

In order to quantify the detectability of individual subhalos (given
our assumptions) we include diffuse Galactic and extragalactic
backgrounds, and convert our photon counts $N_\gamma$ into
significance $S=N_s/\sqrt{N_b}$, where $N_s$ and $N_b$ are the source
and background counts, respectively. For the extragalactic background
we use the EGRET measurement \citep{Sreekumar1998} and for the
Galactic background we follow \citep{Baltz2000} and assume that it is
proportional to the Galactic H I column density
\citep{Dickey1990}. Whereas the extragalactic component is uniform
over the sky, the Galactic background is strongest towards the center
and in a band of $b \pm 10^\circ$ around the Galactic disk.

We consider all objects with $S>5$ to be detectable by GLAST. With our
choice of parameters the halo center could be significantly detected,
with $S>100$. The number of subhalos with $S>5$ depends strongly on
the applied boost factor. Without boosting the subhalo fluxes, only
the most massive halo is detectable. Applying a boost factor of 5
(10), we find that 29 (71) subhalos satisfy the $S>5$ threshold for
detectability. Note that subhalos below our current resolution limit
might also be detectable. Their greater abundance reduces the expected
distance to the nearest neighbor, and this may compensate for their
lower intrinsic luminosities (see Koushiappas' contribution in these
Proceedings).

In conclusion we find that with favorable particle physics parameters,
GLAST may very well detect gamma-ray photons originating from DM
annihilations, either from the Galactic center or from individual
subhalos. This would be a sensational discovery of great importance,
and it is worth including a search for a DM annihilation signal in the
data analysis.

%%%%%%%%%%%%%%%%%%%%%%%%%%%%%%%%%%%%%%%%%%%%%%%%
%% BACKMATTER
%%%%%%%%%%%%%%%%%%%%%%%%%%%%%%%%%%%%%%%%%%%%%%%%

\begin{theacknowledgments}

P.M. acknowledges support from NASA grants NAG5-11513 and NNG04GK85G,
and from the Alexander von Humboldt Foundation. J.D. acknowledges
support from NASA through Hubble Fellowship grant HST-HF-01194.01. The
Via Lactea simulation was performed on NASA's Project Columbia
supercomputer system.

\end{theacknowledgments}


\end{document}